\documentclass[twocolumn,aps,prl,amsmath,amssymb,showpacs,preprintnumbers]{revtex4}

\newcommand{\nc}{\newcommand}

\nc{\be}{\begin{equation}}

\nc{\ee}{\end{equation}}

\nc{\bea}{\begin{eqnarray}}

\nc{\eea}{\end{eqnarray}}

\nc{\xx}{\nonumber\\}

\nc{\ct}{\cite}

\nc{\la}{\label}

\nc{\eq}[1]{(\ref{#1})}

\def\ajou#1&#2(#3){\ \sl#1\bf#2\rm(19#3)}

\def\half{{\frac{1}{2}}}
\def\zbar{{\bar z}}
\def\[{\left [}
\def\]{\right ]}

\begin{document}
\preprint{HU-EP-05/80}
\preprint{hep-th/0512215}
\title{Gravitational Instantons from Gauge Theory}
\author{Hyun Seok Yang and Mario Salizzoni}
\email{hsyang, sali@physik.hu-berlin.de}
\affiliation{Institut f\"ur Physik,
Humboldt Universit\"at zu Berlin,
Newtonstra\ss e 15, D-12489 Berlin, Germany}

\date{\today}

\begin{abstract}
A gauge theory can be formulated on a noncommutative (NC) spacetime. This NC gauge
theory has an equivalent dual description through the so-called Seiberg-Witten
(SW) map in terms of an ordinary gauge theory on a commutative spacetime.
We show that all NC $U(1)$ instantons of Nekrasov-Schwarz type are mapped to
ALE gravitational instantons by the exact SW map and that the NC gauge theory of
$U(1)$ instantons is equivalent to the theory of hyper-K\"ahler geometries.
It implies the remarkable consequence that ALE gravitational instantons can
emerge from local condensates of purely NC photons.
\end{abstract}


\maketitle
It is believed that spacetime must change its nature at distances
comparable to the Planck scale. That is, spacetime at short distances is described by
noncommutative (NC) geometry, where the spacetime coordinates do not commute; order
is important. It may help understanding nonlocality at short distances
since there are reasons to believe that any theory of quantum
gravity will not be local in the conventional sense.

Consider a NC spacetime defined by
\be \la{nc-space}
[y^\mu, y^\nu]_\star = i \theta_{\mu\nu}
\ee
with a constant $4 \times 4$ matrix $\theta_{\mu\nu}$. Gauge theories
can be constructed on this NC spacetime.
For example, the action for NC $U(1)$ gauge theory in flat Euclidean ${\bf R}^4$
is given by
\begin{equation}\label{nced}
\widehat{S}_{\mathrm{NC}} = \frac{1}{4}\int\! d^4 y \,\widehat{F}_{\mu\nu} \star \widehat{F}^{\mu\nu},
\end{equation}
where noncommutative fields are defined by
\be \la{ncf}
\widehat{F}_{\mu\nu}=\partial_{\mu}\widehat{A}_{\nu}-\partial_{\nu}\widehat{A}_{\mu}-
i \,[\widehat{A}_{\mu}, \widehat{A}_{\nu}]_\star.
\ee
It was shown in \ct{nek-sch} that this NC $U(1)$ gauge theory admits non-singular instanton
solutions satisfying the NC self-duality equation,
\be \la{nc-self-dual}
{\widehat F}_{\mu\nu} (y) = \pm \half
\varepsilon_{\mu\nu\alpha\beta}{\widehat F}_{\alpha\beta} (y).
\ee

The NC gauge theory has an equivalent dual description through the so-called Seiberg-Witten
(SW) map in terms of ordinary gauge theory on commutative spacetime \ct{sw}.
The SW map is a map between gauge orbit spaces of commutative and NC gauge
fields. It was shown in \ct{yang,ban-yang} that the commutative nonlinear
electrodynamics equivalent to Eq.\eq{nced} is given by
\begin{equation}\label{ced-sw}
S_{\mathrm{C}} = \frac{1}{4} \int d^4 x \sqrt{\det{{\rm g}}} \;
{\rm g}^{\mu \alpha} {\rm g}^{\beta\nu} F_{\mu\nu}
F_{\alpha\beta},
\end{equation}
where we introduced an ``effective metric" induced
by the dynamical gauge fields such that
\begin{equation}\label{effective-metric}
    {\rm g}_{\mu\nu} = \delta_{\mu\nu} + (F\theta)_{\mu\nu},
    \;\;  ({\rm g}^{-1})^{\mu\nu} \equiv {\rm g}^{\mu\nu} =
    \Bigl(\frac{1}{1 + F\theta}\Bigr)^{\mu\nu}.
\end{equation}
The commutative action \eq{ced-sw} can actually be derived
from the NC action \eq{nced} using the exact SW maps in \ct{ban-yang}
(see \ct{liu} for the exact inverse SW map):
\bea \label{eswmap}
    &&  \widehat{F}_{\mu\nu}(y) = \Bigl(\frac{1}{1 + F\theta} F
    \Bigr)_{\mu\nu}(x), \\
    \la{measure-sw}
    && d^{4} y = d^{4} x \sqrt{\det(1+ F \theta)}(x),
\eea
where $ x^\mu(y) \equiv y^\mu + \theta^{\mu\nu} \widehat{A}_\nu(y).$
It was shown in \ct{sty} that the self-duality equation for the action
$S_{\mathrm{C}}$ is given by
\be \la{c-self-dual}
{\bf F}_{\mu\nu} (x) = \pm \half
\varepsilon_{\mu\nu\alpha\beta}{\bf F}_{\alpha\beta} (x),
\ee
where
\begin{equation}\label{def-fatf}
{\bf F}_{\mu\nu} (x) \equiv \Bigl({\rm g}^{-1} F \Bigr)_{\mu\nu} (x).
\end{equation}
The above equation is directly obtained by the exact SW map \eq{eswmap}
from the NC self-duality equation \eq{nc-self-dual}.
It was checked in \ct{sty} that the field configuration that
satisfies the self-duality equation \eq{c-self-dual} also satisfies the
equations of motion derived from the action \eq{ced-sw}.

A general strategy was suggested in \ct{sty} to solve the self-duality
equation \eq{c-self-dual}. For example, let us
consider the anti-self-dual (ASD) case.
Take a general ansatz for the ASD $ {\bf F}_{\mu\nu}$ as follows
\be \la{bff-ansatz}
{\bf F}_{\mu\nu}(x) = f^a(x)\bar{\eta}^a_{\mu\nu},
\ee
where $\bar{\eta}^a$ are three $ 4 \times 4$ ASD 't Hooft matrices and $f^a$'s are arbitrary
functions. Then the equation \eq{c-self-dual} is automatically satisfied.
Next, solve the field strength $F_{\mu\nu}$ in terms of ${\bf F}_{\mu\nu}$:
\begin{equation}\label{f-fatf}
    F_{\mu\nu} (x) = \Bigl(\frac{1}{1-{\bf F}\theta}{\bf F}\Bigr)_{\mu\nu}(x),
\end{equation}
and impose the Bianchi identity for $F_{\mu\nu}$,
\be \la{bianchi}
\varepsilon_{\mu\nu\rho\sigma}\partial_{\nu} F_{\rho\sigma} = 0,
\ee
since the field strength $F_{\mu\nu}$
is given by a (locally) exact two-form, i.e., $F=dA$. In the end
one can get general differential equations governing $U(1)$ instantons \ct{sty}.

Now let us restrict to the self-dual (SD) NC ${\bf R}^4$ properly normalized as
$\theta_{\mu\nu}= \half \eta^3_{\mu\nu}$ to consider the Nekrasov-Schwarz
instantons \ct{nek-sch}. The NC parameter $\theta$ can be easily recovered
by a simple dimensional analysis by recalling that $\theta$
carries the dimension of $(length)^2$. It was shown \ct{sty} that
the parameter $\theta$, which settles the size of NC instantons,
is related to the size of a minimal two-sphere known as a ``Bolt"
in the gravitational instantons.

Substituting the ansatz \eq{bff-ansatz} into Eq.\eq{f-fatf}, we get
\be \la{field-f}
F_{\mu\nu} = \frac{1}{1-\phi}f^a \bar{\eta}^a_{\mu\nu} -
\frac{2\phi}{1-\phi}\eta^3_{\mu\nu},
\ee
where $\phi \equiv \frac{1}{4}\sum_{a=1}^3 f^a(x)f^a (x).$
From Eq.\eq{field-f}, we also obtain
\be \la{sw-instanton}
F_{\mu\nu}^+ \equiv \half(F_{\mu\nu}+ \half \varepsilon_{\mu\nu\rho\sigma}
F_{\rho\sigma}) = \frac{1}{4}(F\widetilde{F}) \theta_{\mu\nu}^+
\ee
since
\be \la{fdualf}
F\widetilde{F} \equiv \half
\varepsilon^{\mu\nu\rho\sigma}F_{\mu\nu}F_{\rho\sigma}=
- \frac{16 \phi}{1-\phi}.
\ee
We thus get
\bea \la{f=f}
&& F_{24} = F_{13}, \quad  F_{14} = - F_{23}, \\
\la{f12f34}
&& F_{12} + F_{34} = \frac{1}{4} F\widetilde{F}.
\eea
Note that Eq.\eq{sw-instanton} is equivalent to the instanton equation in
\ct{sw}, although there it was derived perturbatively.

The metric for the $U(1)$ fields in Eq.\eq{field-f} becomes symmetric, i.e.,
${\rm g}_{\mu\nu} = {\rm g}_{\nu\mu}$ and its components are
\bea \la{metric1234}
&& {\rm g}_{11} = {\rm g}_{22} = 1-\half F_{12},
\quad  {\rm g}_{33} = {\rm g}_{44} = 1 - \half F_{34}, \xx
&& {\rm g}_{13} = {\rm g}_{24} = - \half F_{14},
\quad  {\rm g}_{14} = - {\rm g}_{23} = \half F_{13}, \\
&& {\rm g}_{12} = {\rm g}_{34} = 0. \nonumber
\eea
Eq.\eq{f12f34} can be rewritten using the metric ${\rm g}_{\mu\nu}$
as
\be \la{real-ma-equation}
{\rm g}_{\mu\mu} = 4 \sqrt{\det{{\rm g}_{\mu\nu}}}
\ee
with $\sqrt{\det{{\rm g}_{\mu\nu}}} = {\rm g}_{11}{\rm g}_{33}-({\rm g}_{13}^2 +
{\rm g}_{14}^2).$
We will show later that Eq.\eq{real-ma-equation} reduces to the so-called
complex Monge-Amp\`ere equation, which is the Einstein field equation for a
K\"ahler metric \ct{kaehler}.

The equation \eq{bianchi} was solved in \ct{sty} for the single instanton case.
It was found there that the effective metric \eq{effective-metric} for the
single $U(1)$ instanton is related to the Eguchi-Hanson (EH) metric \ct{eh}, the
simplest ALE space, and that the family of the EH space is parameterized by the
instanton number. In this paper we will show that the connection between NC
$U(1)$ instantons and hyper-K\"ahler geometries is more general. More
precisely, we will see that the NC self-duality equation \eq{nc-self-dual} is
mapped by the SW map to gravitational instantons, defined by
the following self-duality equation \ct{hawking,egh-pr}
\be \la{instanton-gravity}
R_{abcd} = \pm \half \varepsilon_{abef}
{R^{ef}}_{cd},
\ee
where $R_{abcd}$ is a curvature tensor.

To proceed with the K\"ahler geometry, let us introduce the complex
coordinates and the complex gauge fields
\bea \la{complex-r4}
&& z_1 = x^2 + i x^1, \qquad z_2 = x^4 + i x^3, \\
\la{complex-a}
&& A_{z_1} = A^2 - i A^1, \quad A_{z_2} = A^4 - i A^3.
\eea
In terms of these variables, Eqs.\eq{f=f} and \eq{f12f34} are written as
\bea \la{complexf=f}
&& F_{z_1 z_2}  = 0 = F_{\zbar_1 \zbar_2}, \\
\la{complexf12f34}
&& F_{z_1 \zbar_1} + F_{z_2 \zbar_2 } = - \frac{i}{4} F\widetilde{F},
\eea
where $F\widetilde{F} = -4 ( F_{z_1 \zbar_1} F_{z_2 \zbar_2 }
+  F_{z_1 \zbar_2} F_{\zbar_1 z_2}).$
Using the metric components in Eq.\eq{metric1234}, one can easily see that the
metric ${\rm g}_{\mu\nu}$ is a Hermitian metric. That is,
\be \la{hermitian-metric}
ds^2 = {\rm g}_{\mu\nu} dx^\mu dx^\nu = g_{i\bar{j}} dz_i
d\bar{z}_j, \quad i,j=1,2,
\ee
where
\bea \la{hermitian-metric12}
&& g_{z_1 \zbar_1} =  {\rm g}_{11} = \frac{1}{1-\phi}\Big(1-\frac{f^3}{2}\Bigr), \xx
&& g_{z_2 \zbar_2} =  {\rm g}_{33} = \frac{1}{1-\phi}\Bigl(1 + \frac{f^3}{2}\Bigr), \xx
&& g_{z_1 \zbar_2} =  {\rm g}_{13}  -i {\rm g}_{14}  =
\frac{1}{1-\phi}\frac{f^1+if^2}{2},\\
&& g_{\zbar_1 z_2} =  {\rm g}_{13}  + i {\rm g}_{14}  =
\frac{1}{1-\phi}\frac{f^1-if^2}{2}. \nonumber
\eea

If we let
\be \la{kahler-form}
\Omega = \frac{i}{2} g_{i\bar{j}} dz_i \wedge d\bar{z}_j
\ee
be the K\"ahler form, then the K\"ahler condition is $d\Omega = 0$
\ct{egh-pr}, or, for all $i,j,k$,
\be \la{kahler-condition}
\frac{\partial g_{i\bar{j}}}{\partial z^k} = \frac{\partial g_{k\bar{j}}}{\partial z^i}.
\ee
It is straightforward to check the pleasant property that the K\"ahler
condition \eq{kahler-condition} is equivalent to the Bianchi identity
\eq{bianchi} for the $U(1)$ field strength \eq{field-f}.
Thus our metric $g_{i\bar{j}}$ is a
K\"ahler metric and thus we can introduce a K\"ahler potential $K$
defined by
\be \la{kahler-potential}
g_{i\bar{j}} = \frac{\partial^2 K}{\partial z^i \partial \zbar^j}.
\ee

Now we will show that the K\"ahler potential $K$ is related to the
integrability condition of the self-duality equation
\eq{c-self-dual}. Locally, there is no difficulty in finding the general
solution of Eq.\eq{complexf=f}:
\be \la{complex-gauge}
A_{z_i} = 0, \quad A_{\zbar_i} = 2i \partial_{\zbar_i} (K - \zbar_k z_k).
\ee
Then one can easily check that the real-valued smooth function $K$
in Eq.\eq{complex-gauge} is equivalent to the K\"ahler potential in
Eq.\eq{kahler-potential} (up to holomorphic diffeomorphisms).

As announced in Eq.\eq{instanton-gravity}, the metric ${\rm g}_{\mu\nu}$ in
Eq.\eq{effective-metric} is related to gravitational instantons satisfying
Eq.\eq{instanton-gravity}, whose metrics are necessarily Ricci-flat.
To see this, let us rewrite ${\rm g}_{\mu\nu}$ as
\be \la{new-metric}
{\rm g}_{\mu\nu}= \half(\delta_{\mu\nu} + \widetilde{{\rm g}}_{\mu\nu}).
\ee
Then, from Eq.\eq{real-ma-equation}, one can see that
\be \la{determinant=1}
 \sqrt{\det{\widetilde{{\rm g}}_{\mu\nu}}} = 1.
\ee
Note that the metric $\widetilde{{\rm g}}_{\mu\nu}$ is also a K\"ahler metric:
\be \la{new-kahler}
\widetilde{g}_{i\bar{j}} = \frac{\partial^2 \widetilde{K}}{\partial z^i \partial \zbar^j}.
\ee
The relation $\det{\widetilde{{\rm g}}_{\mu\nu}} = (\det{\widetilde{g}_{i\bar{j}}})^2$
definitely leads to
\be \la{monge-ampere}
\det{\widetilde{g}_{i\bar{j}}} = 1.
\ee
Therefore the metric $\widetilde{{\rm g}}_{\mu\nu}$ is both Ricci-flat and
K\"ahler, which is the case of gravitational instantons \ct{kaehler}.
For example, if one assumes
that $\widetilde{K}$ in Eq.\eq{new-kahler} is a function solely of $r^2 = |z_1|^2 + |z_2|^2$,
Eq.\eq{monge-ampere} can be integrated to give \ct{gibb-pope}
\be \la{pot-k}
\widetilde{K} = \sqrt{r^4 + t^4} + t^2 \log \frac{r^2}{\sqrt{r^4 + t^4} + t^2}.
\ee
This leads precisely to the EH metric of \ct{sty}. The instanton equation \eq{sw-instanton}
is thus equivalent to the Einstein field equation for K\"ahler metrics.

Although it is obvious that the metric $\widetilde{{\rm g}}_{\mu\nu}$ in
Eq.\eq{new-metric} is hyper-K\"ahler \ct{hyper-Kahler}, since in four dimensions the
hyper-K\"ahler condition is equivalent to Ricci-flat K\"ahler, we would like
to show more directly the hyper-K\"ahler structure. First we introduce a dual
basis of 1-forms defined by $\sigma_\mu = \alpha_{\mu\nu} dx^\nu$
where, $\alpha_{\mu\nu} = \sqrt{\gamma {\rm g}_{\mu\nu}}$ for $\mu =\nu$,
 ${\rm g}_{\mu\nu}/2\sqrt{\gamma {\rm g}_{11}}$
for $\mu \neq \nu$ and $\mu = 1,2$, and ${\rm g}_{\mu\nu}/2\sqrt{\gamma {\rm g}_{33}}$
for $\mu \neq \nu$ and $\mu = 3,4$, and
\be
\gamma = \half \Biggl(1 \pm \frac{(\det{{\rm g}_{\mu\nu}})^{\frac{1}{4}}}{\sqrt{{\rm
    g}_{11}{\rm g}_{33}}}\Biggr). \nonumber
\ee
Using the metric components in Eq.\eq{metric1234}, one can show that
the metric can be written as
\be \la{dia-metric}
ds^2 = {\rm g}_{\mu\nu} dx^\mu dx^\nu = \sigma_\mu \otimes \sigma_\mu
\ee
and
\be \la{volume-form}
\sigma_1 \wedge \sigma_2 \wedge \sigma_3 \wedge \sigma_4 = \sqrt{\det{{\rm
      g}_{\mu\nu}}} d^4 x.
\ee

Let us introduce a SD local triple of 2-forms defined by
\be \la{2-form}
\omega^a = \half \eta^a_{\mu\nu} \sigma^\mu \wedge \sigma^\nu,
\ee
where ${\eta}^a$ are three $ 4 \times 4$ SD 't Hooft matrices. The explicit forms
of $\omega^a$ are given by
\bea \la{explicit-2-form1}
&& \omega^1 = (\det{{\rm g}_{\mu\nu}})^{\frac{1}{4}}(dx^1 \wedge dx^4 + dx^2
\wedge dx^3), \\
&& \omega^2 = (\det{{\rm g}_{\mu\nu}})^{\frac{1}{4}}(dx^2 \wedge dx^4 + dx^3
\wedge dx^1), \\
&&  \omega^3 = {\rm g}_{11} dx^1 \wedge dx^2 + {\rm g}_{33} dx^3 \wedge dx^4
+ \xx
&& \qquad \;\; {\rm g}_{13}(dx^1 \wedge dx^4 + dx^3 \wedge dx^2) +  \xx
&& \qquad \;\; {\rm g}_{14}(dx^3 \wedge dx^1 + dx^4 \wedge dx^2).
\eea
One can easily check that $d\omega^3 = 0$ if and only if the Bianchi identity
\eq{bianchi} is satisfied. Since $\omega^3 = - \Omega$ in Eq.\eq{kahler-form},
this result indeed reproduces the K\"ahler condition \eq{kahler-condition}.
Thus the metric ${\rm g}_{\mu\nu}$ is K\"ahler, as was shown before, but not
hyper-K\"ahler, since  $d\omega^1 = d\omega^2 = 0$ requires $ \det{{\rm
    g}_{\mu\nu}} = {\rm constant}$.

However, if we consider the triple of K\"ahler forms of the metric
$\widetilde{{\rm g}}_{\mu\nu}$ as follows,
\be \la{unit-2-form}
\widetilde{\omega}^a = \half \eta^a_{\mu\nu} \widetilde{\sigma}^\mu
\wedge \widetilde{\sigma}^\nu,
\ee
where $\widetilde{\sigma}^\mu$ are defined in the same way as the $\sigma^\mu$s,
but with ${\rm g}_{\mu\nu}$ replaced by $\widetilde{{\rm g}}_{\mu\nu}$,
we immediately get
\be \la{hyper-kahler-condition}
d\widetilde{\omega}^a = 0, \;\; \forall a
\ee
since $\det{\widetilde{{\rm g}}_{\mu\nu}} = 1$. Thus the metric
$\widetilde{{\rm g}}_{\mu\nu}$ should be a hyper-K\"ahler metric \ct{hyper-Kahler}.
Therefore the hyper-K\"ahler condition has one-to-one correspondence
with the self-duality equation of NC $U(1)$ instantons through the
SW map.

Using this hyper-K\"ahler structure, we can easily understand the
ALF such as the Taub-NUT metric \ct{hawking} as well as ALE
instantons \ct{eh,gibb-hawk}, which are a general class of self-dual,
Ricci-flat metrics with a triholomorphic $U(1)$ symmetry.
The metric is given by
\be \la{gh-metric}
ds^2_{GH} = U^{-1}(dx^4 + \vec{a} \cdot d\vec{x})^2 + U d\vec{x} \cdot d\vec{x},
\ee
where $x^4 $ parameterizes circles and $\vec{x} \in {\bf R}^3$. Since the above
mentioned triholomorphic $U(1)$ symmetry is generated by the Killing field
$\partial/\partial x^4$, the $U(1)$ invariant function $U=U(\vec{x})$ does not depend on
$x^4$, and has to satisfy the condition
\be \la{gh-monopole}
\nabla U = \nabla \times \vec{a}.
\ee
It turns out that the condition \eq{gh-monopole} is equivalent to the
hyper-K\"ahler condition \eq{hyper-kahler-condition}.
To see this, let us introduce a 1-form basis as
\bea \la{gh-1-form}
&& \widetilde{\sigma}^i = \sqrt{U} dx^i, \quad (i=1,2,3), \xx
&& \widetilde{\sigma}^4 = \frac{1}{\sqrt{U}}(dx^4 + \vec{a} \cdot d\vec{x}),
\eea
where the metric reads as
\be \la{gh-dia-metric}
ds^2_{GH} = \widetilde{\sigma}_\mu \otimes \widetilde{\sigma}_\mu.
\ee
It is then easy to see that Eq.\eq{gh-monopole} is equivalent
to the hyper-K\"ahler condition \eq{hyper-kahler-condition} for the K\"ahler forms
\eq{unit-2-form} given by the basis \eq{gh-1-form}.

Thus the ALE and ALF spaces are all hyper-K\"ahler manifolds.
But they have very different asymptotic behaviors of curvature tensors:
The ALE spaces fall like $1/r^6$ while the ALF spaces fall like $1/r^3$
\ct{egh-pr}. This suggests that the ALE metrics describe
gravitational ``dipoles'' \ct{eh} while the ALF metrics describe monopoles
(regarded as gravitational dyons) \ct{hawking}.
So ALF instantons may have a rather similar realization in
terms of NC $U(1)$ instantons on ${\bf R}^3 \times S^1$: NC monopoles
\ct{lee-yi}. It seems to be in harmony with the fact that the ALF
metrics approach a flat metric in the asymptotic ${\bf R}^3$ while the ALE
metrics are asymptotically (locally) flat in ${\bf R}^4$.

In conclusion, we showed that NC instantons satisfying the self-duality
equation \eq{nc-self-dual} are all mapped through the SW map to gravitational instantons,
precisely hyper-K\"ahler metrics, satisfying the self-duality equation
\eq{instanton-gravity}. Our result implies that a nontrivial curved spacetime
can emerge from local condensates of purely NC photons without any
graviton. The manifestation of the emergent gravity from NC
gauge theory is {\it not} in conflict with the
Weinberg-Witten theorem \ct{ww}, stating that an interacting graviton cannot
emerge from an ordinary quantum field theory in the same dimensions, since NC
field theories are neither local nor Lorentz invariant. But, recently, it was
found \ct{twist-symm} that NC field theory is invariant under the twisted
Poincar\'e symmetry where the action of generators is now defined by the
twisted coproduct in the deformed Hopf algebras. Although the energy-momentum tensor,
being nonlinear in fields, is in NC field theories a twisted Poincar\'e
invariant, the Weinberg-Witten theorem is still a consequence of the strict Lorentz
invariance.

The connection between NC $U(1)$ instantons and gravitational instantons
addresses several interesting issues. (These two seem to share at least two underlying
mathematical structures: the twistor space and the holomorphic vector bundle,
which may be important to understand why there exists such connection.)
Using the connection, the problem of gravitational instantons can be addressed
from the viewpoint of NC $U(1)$ instantons and vice versa.
For example, the generalized positive action conjecture \ct{positive-action},
gravitational anomalies and their index theorem \ct{anomaly},
and the hyper-K\"ahler quotient construction of ALE spaces \ct{kronh}.

More extended work including the generalization of the emergent
gravity from NC gauge theories beyond the self-dual sector will be
reported elsewhere. We would like to thank Harald Dorn and
Alessandro Torrielli for useful discussions throughout the course
of the work and for reading the manuscript. This work was
supported by the Alexander von Humboldt Foundation (H.S.Y.) and by
DFG under the project SA 1356/1 (M.S.).


\begin{thebibliography}{55}
\bibitem{hyper-Kahler} A hyper-K\"ahler manifold ${\cal M}$ is defined by a
  quadruple $(I,J,K,g)$, where $I,J$ and $K$ are almost complex structures on
  ${\cal M}$ with $IJ=K$, and $g$ is a K\"ahler metric with respect to $I,J$
  and $K$. The K\"ahler forms on ${\cal M}$ are defined by $\omega^1(X,Y) =
  g(IX,Y), \;\omega^2(X,Y) = g(JX,Y)$ and $\omega^3(X,Y) = g(KX,Y)$.
Then the hyper-K\"ahler structure is equivalent to the condition that
the almost complex structures $I,J$ and $K$ are integrable, i.e.,
$d\omega^1 = d\omega^2 = d\omega^3=0$.
\end{thebibliography}
\end{document}